\documentclass{elsart}
\usepackage{amssymb}

\usepackage{amsfonts}
\usepackage{amsmath}

\input{tcilatex}

\begin{document}

\begin{frontmatter}

\title{Quantum Mechanical Ground State of Hydrogen Obtained from Classical Electrodynamics}

\author{Daniel C. Cole and Yi Zou}

\address{Dept. of Manufacturing Engineering, 15 St. Mary's Street, Boston University, Brookline, MA 02446}

\begin{abstract}
The behavior of a classical charged point particle under the influence of only a
Coulombic binding potential and classical electromagnetic zero-point radiation,
is shown to yield agreement with the probability density distribution of Schrödinger's
wave equation for the ground state of hydrogen. These results, obtained without
any fitting parameters, again raise the possibility that the main tenets of
stochastic electrodynamics (SED) are correct, thereby potentially providing
a more fundamental basis of quantum mechanics.  The present methods should
help propel yet deeper investigations into SED.
\end{abstract}


\end{frontmatter}

The following fact probably comes as a surprise to most physicists. \ A
group of researchers in the past have both proposed and deeply investigated
the idea that classical electrodynamics, namely, Maxwell's equations and the
relativistic version of Newton's equation of motion, may describe much, if
not all, of atomic physical processes, provided one takes into account the
appropriate classical electromagnetic random radiation fields acting on
classical charged particles. \ Stochastic electrodynamics (SED) is the usual
name given for this physical theory; it was most significantly advanced in
the 1960s by Boyer \cite{Boyer1969Black1},\cite{Boyer1969Black2} and
Marshall \cite{Marshall1963}, \cite{Marshall1965a}, \cite{Marshall1965b},
although its full history is somewhat more complicated and is reviewed and
described in Ref. \cite{delaPena1996}. \ Other useful reviews exist such as
Refs. \cite{Boyer1980review}, \cite{Boyer1975review1}, and \cite%
{Cole1993reviewSP}. \ 

SED is really a subset of classical electrodynamics. \ However, it differs
from conventional treatments in classical electrodynamics in that it assumes
that if thermodynamic equilibrium of classical charged particles is at all
possible, then a thermodynamic radiation spectrum must also exist and must
be an essential part of the thermodynamic system of charged particles and
radiation. \ As can be shown via statistical and thermodynamic analyses \cite%
{Boyer1969Black1}, \cite{Cole19901st}, if thermodynamic equilibrium is
possible for such a system, then there must exist random radiation that is
present even at temperature $T=0$. \ This radiation has been termed
classical electromagnetic zero-point (ZP) radiation, where the
\textquotedblleft ZP\textquotedblright\ terminology stands for $T=0$, as
opposed to \textquotedblleft ground state\textquotedblright\ or
\textquotedblleft lowest energy state\textquotedblright. \ Either of the
following requirements has been shown to enable the derivation of the
required functional form of the ZP radiation spectrum: (1) the ZP radiation
must possess a Lorentz invariant character \cite{Boyer1969Black1}, and (2)\
no heat must flow during reversible thermodynamic operations \cite%
{Cole19901st},\cite{Cole19902nd},\cite{Cole19921st}. \ Deriving the ZP
spectral form from (1) follows only from the radiation properties, while (2)
involves the interaction of both particles and fields.

Results have been obtained from SED that agree nicely with quantum
mechanical (QM) predictions for linear systems \cite{Boyer1980review}, such
as for systems of electric dipole simple harmonic oscillators \cite%
{Boyer1975review2},\cite{Cole1993reviewSP}, and for linear electromagnetic
fields in Casimir/van der Waals type situations \cite{delaPena1996},\cite%
{Cole19921st},\cite{Cole2000ConductCav}. \ Moreover, most physicists, who
know of SED, are likely to agree that SED provides a better description of
physical processes than does conventional classical electrodynamics without
the consideration of ZP and Planckian electromagnetic radiation. \
Nevertheless, since the late 1970s and early 1980s, the vast majority of
physicists have clearly concluded that SED cannot come close to predicting
the full range of QM phenomena for nonlinear dynamics found in real atomic
systems \cite{MarshallHydrogen1980}, \cite{ClaverieHydrogen1980}, \cite%
{ClaverieHydrogen1981},\cite{ClaverieHydrogen1982}, \cite{blanco1}, \cite%
{blanco2}, \cite{Cole1993reviewSP}, \cite{delaPena1996}. \ In particular,
these past analyses of SED predicted clear disagreements with physical
observation, such as that a single hydrogen atom will ionize at $T=0$ and
that the spectra predicted by SED does not agree with QM predictions.

However, as discussed in Refs. \cite{Boyer1989} and \cite{Cole1990FoundPhys}%
, reasons exist to raise some doubts on these conclusions. \ In particular,
for atomic systems, all of the key physical effects should arise from
electromagnetic interactions. \ Examining other nonlinear binding
potentials, other than ones arising from Coulombic binding potentials, have
no relation to real physical atomic systems. \ Even though one can place any
potential function in Schr\"{o}dinger's equation, and attempt to solve it,
SED does not need to match these solutions as they have little relationship,
in detail, to the real physical world of atomic systems. \ Instead,
realistic binding potentials must be examined. \ Moreover, for perturbation
analyses, if one assumes that the small effect of the electric charge is a
key part of the perturbation analysis, then this effect must be consistently
carried out for the radiation reaction as well as for the binding potential
and the effect of the ZP field acting on the orbiting charge \cite%
{Cole1990FoundPhys}. \ Still, properly accounting for these objections into
an improved analytic, or even semi-analytic, reanalysis of SED, has seemed
quite difficult.

For that reason, in this article we have turned to attacking one of the more
significant problems in SED via simulation methods, namely, the hydrogen
atom. \ The present results certainly seem to bear out the hope that the
earlier impasse in SED may have been due to the difficulties of analyzing
nonlinear stochastic differential equations, rather than a fundamental
physical flaw in the basic ideas of SED.

In quick summary, the present simulation work was carried out by tracking\
individual trajectories of electrons for long lengths of time, assuming
classical electrodynamics governed the trajectories. \ Probability
distributions were then obtained in coordinate space based on the length of
time the electrons spent in regions of space about the nucleus. \ References 
\cite{ColeWilson2002},\cite{ColeWilson2nd2002}, \cite{ColeWilson3rd2003},
and \cite{ColeWilson4th2003} contain many of the technical details that led
to the present work, although these previous works concentrated on the
nonlinear dynamical effects of a classical electron, with charge $-e$ and
rest mass $m$, in orbit about an infinitely massive nucleus of charge $+e$,
where besides the binding potential acting, only a limited set of plane
waves acted on the electron. \ In that work, as here, we have numerically
solved the nonrelativistic approximation to the classical Lorentz-Dirac
equation \cite{dirac},\cite{teitelboim}:%
\begin{equation*}
m\mathbf{\ddot{z}}=-\frac{e^{2}\mathbf{z}}{\left\vert \mathbf{z}\right\vert
^{3}}+\mathbf{R}_{\text{reac}}+\left( -e\right) \left\{ \mathbf{E}\left[ 
\mathbf{z}\left( t\right) ,t\right] +\frac{\mathbf{\dot{z}}}{c}\times\mathbf{%
B}\left[ \mathbf{z}\left( t\right) ,t\right] \right\} \ \ , 
\end{equation*}
where the radiation reaction term of $\mathbf{R}_{\text{reac}}$ has been
approximated by $\mathbf{R}_{\text{reac}}\thickapprox\frac{2}{3}\frac{e^{2}}{%
c^{3}}\frac{d^{3}\mathbf{z}}{dt^{3}}\thickapprox\frac{2}{3}\frac{e^{2}}{c^{3}%
}\frac{d}{dt}\left( -\frac{e^{2}\mathbf{z}}{m\left\vert \mathbf{z}%
\right\vert ^{3}}\right) $, and where $\mathbf{E}$ and $\mathbf{B}$
represent the electric and magnetic fields of the radiation acting on the
electron. \ We note that to date we have carried out a fair bit of numerical
analysis involving full relativistic computation, but, for the results
reported here, the key effects of our present system are adequately
represented by the above equations.

The electromagnetic ZP field formally consists of an infinite set of
frequencies, which clearly would be impossible to implement fully in any
sort of numerical scheme. \ Consequently, we limited the number of
frequencies in the simulation to ranges that had the most significant effect
on the electron's orbital motion. \ We did so in two ways. \ Often the ZP
radiation fields are represented in SED by a sum of plane waves \cite%
{delaPena1996}:%
\begin{equation*}
\mathbf{E}_{\text{ZP}}\left( \mathbf{x},t\right) =\frac{1}{\left(
L_{x}L_{y}L_{z}\right) ^{1/2}}\sum\limits_{n_{x},n_{y},n_{z}=-\infty
}^{\infty }\sum\limits_{\lambda =1,2}\mathbf{\hat{\varepsilon}}_{\mathbf{k}_{%
\mathbf{n}},\lambda }\left[ 
\begin{array}{c}
A_{\mathbf{k}_{\mathbf{n}},\lambda }\cos \left( \mathbf{k}_{\mathbf{n}}\cdot 
\mathbf{x-}\omega _{\mathbf{n}}t\right)  \\ 
+B_{\mathbf{k}_{\mathbf{n}},\lambda }\sin \left( \mathbf{k}_{\mathbf{n}%
}\cdot \mathbf{x-}\omega _{\mathbf{n}}t\right) 
\end{array}%
\right] \ \ ,
\end{equation*}%
with $n_{x}$, $n_{y}$, and $n_{z}$ integers, $\mathbf{k}_{\mathbf{n}}=2\pi
\left( \frac{n_{x}}{L_{x}}\mathbf{\hat{x}+}\frac{n_{y}}{L_{y}}\mathbf{\hat{y}%
+}\frac{n_{z}}{L_{z}}\mathbf{\hat{z}}\right) $,$\omega _{\mathbf{n}%
}=c\left\vert \mathbf{k}_{\mathbf{n}}\right\vert $, $\mathbf{k}_{\mathbf{n}%
}\cdot \mathbf{\hat{\varepsilon}}_{\mathbf{k}_{\mathbf{n}},\lambda }=0$, $%
\mathbf{\hat{\varepsilon}}_{\mathbf{k}_{\mathbf{n}},\lambda }\cdot \mathbf{%
\hat{\varepsilon}}_{\mathbf{k}_{\mathbf{n}},\lambda ^{\prime }}=0$ for $%
\lambda \neq \lambda ^{\prime }$, and $A_{\mathbf{k}_{\mathbf{n}},\lambda }$
and $B_{\mathbf{k}_{\mathbf{n}},\lambda }$ are both real quantities. \ $%
\mathbf{B}_{\text{ZP}}\left( \mathbf{x},t\right) $ is expressed by replacing 
$\mathbf{\hat{\varepsilon}}_{\mathbf{k}_{\mathbf{n}},\lambda }$ by $\left( 
\mathbf{\hat{k}}_{\mathbf{n}}\mathbf{\times \hat{\varepsilon}}_{\mathbf{k}_{%
\mathbf{n}},\lambda }\right) $ in the above expression for $\mathbf{E}_{%
\text{ZP}}\left( \mathbf{x},t\right) $. \ In the above, $L_{x}$, $L_{y}$,
and $L_{z}$ are dimensions of a rectilinear region in space. \ Usually at
the end of SED calculations, these dimensions are taken to a limit of
infinity. \ For our simulation, we wanted them to be large, but not so large
that they created too many plane waves to prohibit numerical simulation. \
The coefficients $A_{\mathbf{k}_{\mathbf{n}},\lambda }$ and $B_{\mathbf{k}_{%
\mathbf{n}},\lambda }$ were taken to be independent random variables
generated once at the start of each simulation, via a random number
generator routine, and then held fixed in value for the remainder of the
simulation. \ The random number generator algorithm was designed to produce
a Gaussian distribution for these coefficients, with an expectation value of
zero, and a second moment of, $\left\langle A_{\mathbf{k}_{\mathbf{n}%
},\lambda }^{2}\right\rangle =\left\langle B_{\mathbf{k}_{\mathbf{n}%
},\lambda }^{2}\right\rangle =2\pi \hbar \omega _{\mathbf{n}}$. \ The latter
specification corresponds to the energy spectrum of classical
electromagnetic ZP radiation of $\rho _{\text{ZP}}\left( \omega \right)
=\hbar \omega ^{3}/\left( 2\pi c^{3}\right) \,$\cite{delaPena1996}.

For reasons to be explained shortly, the orbit of the electron was forced to
lie in the $x-y$ plane. \ We retained plane waves in our simulation from the
summation expression above for the ZP fields, up to an angular frequency
that corresponded to that of an electron in a circular orbit of radius $0.1%
\unit{\mathring{A}}$, or, $\omega_{\text{max}}\approx5.03\times10^{17}\unit{s%
}^{-1}$. For our simulations, we chose $L_{x}=L_{y}=37.4\unit{\mathring{A}}$
and $L_{z}=40,850,000\unit{\mathring{A}}\approx0.41\unit{cm}$, bearing in
mind that this scenario has some similarity to an atom situated in a
rectilinear cavity with highly conducting walls of these dimensions; thus,
this \textquotedblleft cavity\textquotedblright, or region of space, was
made very narrow ($\approx37\unit{\mathring{A}}$), but still fairly large in
width compared to the Bohr radius ($\approx0.53\unit{\mathring{A}}$), and
comparatively very long ($\approx0.41\unit{cm}$). \ This procedure was done
to keep the number of plane waves needed as small as possible, while still
attempting to retain the most important physical effects. \ By making $L_{x}$
and $L_{y}$ so very much smaller than $L_{z}$, then if $n_{x}$ or $n_{y}$
was anything other than zero, the frequency of the associated plane wave
would be greater than $c2\pi /L_{x}\approx5.04\times10^{17}\unit{s}^{-1}$,
thereby enabling us to drop such waves in this approximation scheme. \
Consequently, only waves traveling in the $+\mathbf{\hat{z}}$ and $-\mathbf{%
\hat{z}}$ were retained; the value of $L_{z}$ we chose then resulted in $%
\approx2.2\times10^{6}$ plane being used in the simulation. \ The minimum,
nonzero, angular frequency in the simulation was $\omega_{\text{min}%
}=c2\pi/L_{z}\approx4.61\times10^{11}\unit{s}^{-1}$, which corresponds to
the angular frequency of an electron in a circular orbit of\ radius $%
\approx1.06\times10^{-5}\unit{cm}$, or, about 2000 times the size of the
Bohr radius, $a_{\text{B}}\approx0.53\unit{\mathring{A}}$. \ In this way, we
expected to simulate the approximate behavior of the classical electron in
the SED scheme, for radii lying between about $0.1\unit{\mathring{A}}$ to
hundreds of Angstroms.

This approximate method for representing the desired physical situation
greatly reduced the number of plane waves required if $L_{x}$, $L_{y}$, and $%
L_{z}$ were all made equal to $\approx 0.41\unit{cm}$. \ Although physically
this last approach would be more desirable, it would have resulted in an
absurd number of plane waves to handle numerically, namely, $\left(
2.2\times 10^{6}\right) ^{3}\approx 10^{19}$ waves. \ Nevertheless, even our
much reduced number of $2.2\times 10^{6}$ waves created expensive runs in
CPU time. \ Consequently, we experimented with and found a second
approximation method that reduced our CPU times yet further, while still
retaining key physical effects. \ We will refer to this second method as our
\textquotedblleft window\textquotedblright\ approximation. We note that the
results of our window approximation described below have produced results
that agree reasonable well with other simulation tests we have made that do
not invoke this window approximation, but that require CPU times of about
250 times what we report below.

Specifically, as discussed in Refs. \cite{ColeWilson2002} and \cite%
{ColeWilson4th2003}, we found that each plane wave effected near-circular
orbits most significantly for orbital angular frequencies lying within a
fairly narrow range of the angular frequency of the plane wave itself. \
Figure 9 in Ref. \cite{ColeWilson4th2003} best illustrates this point. \ Our
numerical experiments found that for the average range of plane wave
amplitudes in the present simulation scheme, that a window of $\pm3\%$ about
each average radius more than adequately accounted for the most significant
effects. \ We were prepared to examine a much more complicated window
algorithm due to elliptical orbit considerations, based on the work of Ref. 
\cite{ColeWilson2nd2002}, but numerical experiments showed that the
eccentricity of the orbits remained small throughout the simulation runs,
thereby reducing the need for such considerations. \ Since the angular
frequency of the classical electron in a circular orbit is $e/\left(
mr^{3}\right) ^{1/2}$, the specific algorithm we implemented kept track of
the radius $r$ and retained in the simulation the plane waves with angular
frequencies that fell within a range of $e/\left( mr_{H}^{3}\right) ^{1/2}$
to $e/\left( mr_{L}^{3}\right) ^{1/2}$, where $r_{L}=r\left( 1-f\right) $
and $r_{H}=r\left( 1+f\right) $, where $f$ was selected in these simulations
to be 0.03, based on resonance width analysis. \ As $r$ changed, this scheme
automatically changed the range of plane wave frequencies included in the
summation to act on the electron, but always considered only those specific
plane waves already initialized via the random number generation carried out
at the beginning of the simulation. \ Future speedups in the simulation
might well profit by lowering the value of $f$ yet further, and/or by
treating it as a function of $r$ to better fit resonance width as $r$ varies.

A typical simulation produced roughly circular orbits that would grow and
shrink in radius over time, as seen in Fig. 1. \ We carried out 11
simulations, each with the starting condition of $r=0.53\unit{\mathring{A}}$%
, but with different seeds in the random number generation scheme to create
a different set of plane waves. \ Consequently, the trajectory of each of
these simulations was completely different, although the general character
of each was similar. \ We used a Runge-Kutta 5th order algorithm, with an
adaptive stepsize. \ The simulation code was written in C; the runs were
carried out on 11 separate Pentium 4 PCs, each with 1.8\textbf{\ }GHz
processing speed and 512 MB of RAM. \ The CPU times for each run was about 5
CPU days, with some more and some less, as we attempted to have all
electrons tracked for reasonably close to the same length in time. \
However, for those electrons spending more time near the nucleus, the
calculations took longer because of the faster fluctuations involved. \ The
net time for all runs was about 55 CPU days.

Each of the four snapshots in Fig. 2 show the radial probability density
curve, $P_{\text{QM}}\left( r\right) $ vs. $r$, from Schr\"{o}dinger's wave
equation for the ground state of hydrogen, versus the probability
distribution calculated at the indicated snapshot in time. \ In Fig. 2(a),
the simulated trajectories still strongly show the character of the initial
condition of $r=0.53\unit{\mathring{A}}$. \ However, each succeeding
snapshot shows a striking convergence toward $P_{\text{QM}}\left( r\right) $%
. \ Moreover, the probability distribution for the end of each of the
individual eleven runs has a reasonable resemblance to $P_{\text{QM}}\left(
r\right) $, although combining all of the results together provides a better
match, presumably due to the net longer simulation run and the greater
sampling over field conditions. \ We anticipate that future tests of
interest will involve other initial starting points, deeper testing for
ergodicity, etc.

These simulation results follow the qualitative idea that Boyer originally
suggested in 1975 \cite{Boyer1975review1} that for larger radial orbits, the
dominant part of the ZP spectrum that will effect the orbit will be the low
frequency regime, which has a low energetic contribution, thereby leading on
average to a decaying behavior of the orbit. \ However, for orbits of
smaller radius, then the electron will interact most strongly with the
higher frequency components of the ZP field, which have a larger energetic
contribution. \ Hence, for smaller radii, the probability greatly increases
that the ZP field will act to increase the orbit size. \ In this way, a
stochastic-like pattern should emerge for the electron [Fig. 1].

Without question, the simulations presented here do not \textquotedblleft
prove\textquotedblright\ that SED works for atomic systems. \ There are far
more tests and phenomena to still be examined, including relativistic
corrections and high frequency effects, atomic spectra, many electron
situations, spin, and an understanding of how \textquotedblleft
photon\textquotedblright\ behavior arises. \ We are presently investigating
some of these areas. \ Nevertheless, there is also the very real
possibility, far stronger now that we see predictions for the hydrogen atom
in fairly close agreement with physical observation, that the core ideas of
SED provide a fundamental perspective on nature and a potential basis for QM
phenomena.

\bibliographystyle{elsart-num}
\bibliography{acompat,SED}

\pagebreak

\begin{center}
\textbf{Figure Captions}
\end{center}

Figure 1: Typical plot of $r$ vs. $t$ for one trajectory realization via the
methods described here. \ The inset shows the probability density $P\left(
r\right) $ vs $r$ computed for this particular trajectory.

Figure 2: Plots of the radial probability density vs. radius. \ The solid
line was calculated from the ground state of hydrogen via Schr\"{o}dinger's
equation: $P\left( r\right) =4\pi r^{2}\left\vert \Psi \left( \mathbf{x}%
\right) \right\vert ^{2}=\frac{4r^{2}}{a_{B}^{3}}\exp \left( -\frac{2r}{a_{B}%
}\right) $, where $a_{B}=\hbar ^{2}/me^{2}$. \ The dotted curves are the
simulation results, calculated as a time average for all eleven simulation
runs from time $t=0$ to the average time indicated: (a) $1.417\times 10^{-12}
$ sec; (b) $4.500\times 10^{-12}$ sec; (c) $5.705\times 10^{-12}$ sec; (d) $%
7.252\times 10^{-12}$ sec.

\end{document}